%% file: main.tex
\documentclass[sigconf]{acmart}

\input{_header}

\begin{document}
\setlength{\textfloatsep}{8pt plus 1pt minus 2pt}

\title{Detect, Repair and Audit: A Hierarchical Multi-Agent Defense Against Adversarial Attacks in Energy Forecasting} 

\author{
    {\rm
        Zewei Shi$^{\dagger\ddagger}$,
        Ruoxi Sun$^{\ddagger}$,
        Minhui Xue$^{\ddagger}$,
        Feng Liu$^{\dagger}$,
        Hua Ma $^{\ddagger}$,
        Xingliang Yuan$^{\dagger}$,
        Surya Nepal$^{\ddagger}$
    }\\[0.5em]
    \textit{$^{\dagger}$The University of Melbourne} \quad \textit{$^{\ddagger}$CSIRO} 
}

\input{section/0_abstract}

\maketitle
\pagestyle{plain}

\input{section/1_introduction}
\input{section/2_relatedwork}
\input{section/3_threatmodel}
\input{section/4_methodology}
\input{section/5_experiment_and_evaluation}
\input{section/7_conclusion}
\input{section/999_appendix}

\newpage
\bibliographystyle{ACM-Reference-Format}
\bibliography{references}

\end{document}

%% file: _header.tex
\usepackage[T1]{fontenc}

\usepackage{tikz}
\usepackage{amsmath}
\usepackage{enumitem}

\usepackage{algorithm}       
\usepackage{algpseudocode}
\usepackage{setspace} 
\usepackage{natbib}
\usepackage{graphicx}
\usepackage{soul}
\usepackage{tcolorbox}
\tcbuselibrary{breakable} 

\usepackage{multirow}
\usepackage{booktabs}
\usepackage{makecell}

\def\eg{\emph{e.g.,}\xspace}

\def\ie{\emph{i.e.,}\xspace}
\def\etal{\emph{et al.}\xspace}

\usepackage{framed}
\definecolor{formalshade}{rgb}{0.95,0.95,0.97}
\definecolor{darkblue}{rgb}{0.14,0.22,0.52}
\newenvironment{takeaway}{
  
  \MakeFramed{\advance\hsize-\width\FrameRestore}
  \noindent\hspace{-4.55pt}
}
{
  \endMakeFramed%
}

\usepackage{xcolor}
\usepackage{xspace}

\setcopyright{none}             
\renewcommand\footnotetextcopyrightpermission[1]{}

\usepackage{enumitem}
\setlist[itemize]{itemsep=0pt, parsep=0pt, nosep}

%% file: section/0_abstract.tex
\begin{abstract}
Accurate electricity-demand forecasting is essential for the modern energy infrastructure. Recently, Times-Series Large Language Models (TS-LLMs) have emerged as a promising approach to forecasting more accurate demand from historical consumption data in zero- and few-shot settings, helping energy operators effectively plan supply and allocate energy resources.
However, these models are vulnerable to compromised sensor inputs that can distort forecasts and lead to erroneous market orders or unnecessary load shedding. Existing defenses focus primarily on end-device forecasting, overlooking the distinct data-drift, latency, and audit requirements across energy infrastructure. To address this gap, we extract from major energy standards and regulatory frameworks a three-layer conceptual model that captures the security postures for each layer. Guided by this model, we develop a Multi-Agent Reinforcement Learning (MARL) detector coupled with signal repair at local control centers, designed to mitigate distributional data drift through complementary feature-based detection and clean-reference-calibrated repair while meeting latency budgets. We further deploy an evidence-grounded LLM auditing agent with human-in-the-loop validation for audit-critical Virtual Power Plants (VPPs). In our evaluation, the best MARL configuration achieves an F1 score above 0.91 across all dataset-attack pairs, while signal repair reduces adversarial-to-clean input Relative Squared Error (RSE) by up to 83\%. The few-shot auditor detects 77.78\% of attacks in the Sherlock dataset. On the evaluated hardware, detection and repair require 7.95\,ms on average, with a worst-observed latency of 14.50\,ms, remaining within the local-center latency budget informed by IEEE~1646.
\end{abstract}

%% file: section/1_introduction.tex
\section{Introduction}\label{sec_introduction}
With the rapid proliferation of AI, energy has emerged as a critical foundation for AI infrastructure, powering both the computational resources used to train large language models (LLMs) and the cloud infrastructure that delivers AI inference services to end users~\cite{chen2025electricity,datacenters2025cloudvsedge}. Leading cloud providers are now securing direct nuclear supply agreements~\cite{company-power-planet-evidence-3} and co-locating data centers adjacent to generation assets~\cite{company-power-planet-evidence-1,company-power-planet-evidence-2}.  Within this infrastructure, electricity-demand forecasts derived from historical consumption data inform key decisions on supply scheduling, resource allocation, and market operations~\cite{keep_lights_on}. This growing reliance on energy infrastructure makes its security inseparable from AI service continuity. However, the need to protect this infrastructure is still underscored by real-world incidents. Industroyer malware exploited Operational Technology (OT) protocol automation to trigger the 2016 blackout~\cite{app15179233}, and federal agencies recently warned that state-sponsored actors are actively exploiting programmable logic controllers at US energy and water facilities~\cite{nickel2026iran}. To model defenses across this heterogeneous infrastructure, we synthesize major energy standards and regulatory frameworks covering device communication, grid cybersecurity, and distributed resource aggregation~\cite{etherden2014vpp,nistir7628,ferc2222}. We derive from them a generalized three-layer model (Figure~\ref{fig:framework_overview}): resource-constrained \emph{end devices (L1)}, latency-sensitive \emph{local centers (L2)}, and audit-critical \emph{Virtual Power Plants (VPPs, L3)}, with distinct computational, latency, and audit requirements (as detailed in \S\ref{subsec:infrastructure_goal}).

Traditional energy-demand forecasting relies on historical patterns (\eg daily activity cycles and seasonal variations), while Time-Series Large Language Models (TS-LLMs)~\cite{das2024decoder-timesfm,jin2023timellm-etth2,garza2023timegpt} leverage pre-trained representations for zero- and few-shot forecasting from historical data, making them a more promising demand forecasters at local control centers. However these capabilities do not guarantee robustness to \emph{distributional data drift}: weather or behavioral changes (\eg heatwaves increasing cooling demand) can shift demand away from historical distributions and reduce forecast accuracy. Beyond such benign changes, these models are vulnerable to adversarial perturbations~\cite{zhang2025timeseriesfoundationmodelsdeploymentready}, including black-box attacks that degrade forecasts using only input-output queries without parameter access~\cite{liu2024adversarial,zhang2025timeseriesfoundationmodelsdeploymentready}. Yet existing defenses address only a fragment of this threat landscape. Specifically, adversarial training~\cite{goodfellow2015explainingharnessingadversarialexamples} can be costly for large TS-LLMs and overfit known attack distributions~\cite{rice2020overfitting,yu2022understanding}, while randomized smoothing~\cite{Cohen2019Certified} provides guarantees under specific perturbation models that may not capture temporal attacks on energy data. Anomaly-detection classifiers (\eg Isolation Forest~\cite{liu2008isolation}, OC-SVM~\cite{scholkopf2001estimating}) and signal filters (\eg Kalman smoother~\cite{kalman1960new}, Savitzky-Golay~\cite{savitzky1964smoothing}) must also accommodate benign distribution shifts~\cite{fenza2019drift,lu2019learning} rather than misclassifying or suppressing legitimate changes. Detection-only pipelines that discard suspicious inputs can interrupt forecast continuity, making signal repair an operational necessity. Meanwhile, detection and repair must also share the latency budget with message transmission when performed within control-message delivery paths subject to IEEE~1646's 16\,ms limit~\cite{ieee1646} (\S\ref{sec:threat_model}). Beyond forecasting, VPPs require cross-device OT-log investigation, but manual auditing faces substantial evidence-preparation burdens and gaps in auditors' OT-domain knowledge~\cite{ccs2025-USElectricPowerSector}. Together, these gaps motivate adversarial detection and signal repair that accommodate benign distributional drift within latency constraints, complemented by evidence-grounded auditing at VPPs.

We address these gaps by modeling the energy infrastructure as three layers (\S\ref{sec:threat_model}) and developing a hierarchical defense that combines a cooperative \textit{Multi-Agent Reinforcement Learning (MARL) detector with Discrete Wavelet Transform (DWT)-based signal repair} at local control centers (L2) with an evidence-grounded \textit{LLM auditing agent} at VPPs (L3) (\S\ref{sec:defense}). We take distributional data drift into consideration through two complementary mechanisms. First, the MARL specialists combine evidence from six specialists, allowing the coordinator to assess a load change in context. Training across diverse temporal patterns and conditioning decisions on a rolling trusted context support context-responsive detection (\S\ref{sec:marl}). Second, for rejected inputs, a \textit{Discrete Wavelet Transform (DWT)-based repair module} retrieves a similar clean historical window from the same user and calibrates wavelet soft-thresholds per window. Regarding latency, our experiments show that the full detection-and-repair pipeline achieves a mean latency of 7.95\,ms and a worst observed latency of 14.50\,ms on the evaluated hardware, both below the local-center budget (\S\ref{sec:eval_perf}). At VPPs, the auditing agent correlates OT logs across devices to identify attacks and produces reports grounded in specific log entries for human-in-the-loop (HITL) validation (\S\ref{sec:active}). 

In summary, our contributions are threefold:
\begin{itemize}
    \item We systematically model energy infrastructure as three operational layers (\ie end devices, local control centers, and VPPs) and formalize the threat model for each layer, identifying its data streams, attack surfaces, attacker objectives, and operational constraints (\S\ref{sec:threat_model}).

    \item We develop a hierarchical defense framework combining cooperative MARL detection and DWT-based signal repair at local control centers with evidence-grounded LLM auditing and HITL validation at VPPs, addressing distinct latency and audit requirements while preserving forecast continuity (\S\ref{sec:defense}).

    \item We evaluate local-center defense across three energy datasets under DGA~\cite{liu2024adversarial} and GenIAS~\cite{darban2025geniasgeneratorinstantiatinganomalies}, and separately assess forensic auditing on Sherlock~\cite{Sherlock} benchmark. The best MARL configuration achieves F1 above 0.91 on every dataset--attack pair, DWT repair reduces adversarial-to-clean input RSE by up to 83\%, and the few-shot auditing agent detects 77.78\% of attack windows. On the evaluated hardware, detection and repair take 7.95\,ms on average (14.50\,ms worst observed), below the local-center budget (\S\ref{sec:evaluation}).
\end{itemize}

%% file: section/2_relatedwork.tex
\section{Related Work}
In this section, we review (i) time-series forecasting models and input-side adversarial threats, and (ii) forecast-integrity defenses and audit-governance challenges in energy infrastructure.

\subsection{Time-Series Forecasters and Attacks}
\noindent\textbf{Forecasting models.}
Time-series forecasting has progressed from statistical methods (\eg ARIMA and exponential smoothing) to neural models (\eg DeepAR~\cite{DeepAR} and LSTNet~\cite{lai2018modeling}) and Transformers (\eg Informer~\cite{zhou2021informer} and PatchTST~\cite{Yuqietal-2023-PatchTST}). More recently, TS-LLMs offer zero- and few-shot forecasting capabilities. In this work, we study TimeGPT~\cite{garza2023timegpt}, TimesFM~\cite{das2024decoder-timesfm}, and TimeLLM~\cite{jin2023timellm-etth2} to cover three distinct designs: cross-domain time-series pretraining, decoder-only forecasting, and language-model reprogramming. However, forecasting alone does not ensure robustness to adversarial inputs~\cite{zhang2025timeseriesfoundationmodelsdeploymentready}.

\noindent\textbf{Input-side threats.}
Gradient-based attacks (\eg FGSM/BIM~\cite{goodfellow2015explainingharnessingadversarialexamples} and PGD~\cite{madry2017towards-PGD}) have been extended to time-series classification~\cite{oregi2018adversarial-PGD-TSC} and graph-based forecasting~\cite{liu2022practical}. DGA~\cite{liu2024adversarial} directly targets LLM-based forecasters through gradient-free model queries, producing misleading predictions with small input perturbations. GenIAS~\cite{darban2025geniasgeneratorinstantiatinganomalies}, originally developed for anomaly generation, synthesizes realistic anomalous sequences through latent-space perturbations. These methods perturb time-series inputs; in our setting, corrupting the edge meter's readings can therefore bias local-center forecasts and lead to incorrect VPP market decisions, motivating protection beyond the forecasting model alone.

\subsection{Defenses for Secure Energy Infrastructure}
\noindent\textbf{Forecast integrity.}
Ma~\etal~\cite{keep_lights_on} detect adversarial TS-LLM inputs through \emph{sampling-induced divergence}, checking forecast consistency across shortened input variants. TimeGPT~\cite{garza2023timegpt} offers native anomaly detection, with its historical interface flagging observations outside prediction-based intervals. These detection mechanisms provide anomaly signals but do not specify how to repair adversarially perturbed inputs for downstream forecasting.

\noindent\textbf{Audit governance.}
Sahin~\etal~\cite{ccs2025-USElectricPowerSector} study NERC-CIP compliance in the US electric power sector, identifying heavy evidence-preparation burdens, gaps in auditors' OT-domain knowledge, and a one-size-fits-all audit process. Their study characterizes governance and audit-practice challenges rather than an algorithmic mechanism for detecting OT attacks from device logs. These findings motivate evidence-grounded forensic assistance with human validation.

\noindent\textbf{Cross-layer defense gap.}
These works address complementary security needs but do not jointly provide detection, input recovery, and forensic auditing under distributional data drift and layer-specific operational constraints. Our framework combines cooperative MARL detection with DWT-based repair at local control centers to preserve forecast continuity within their latency budgets, alongside evidence-grounded LLM auditing agent and human validation at VPPs.

%% file: section/3_threatmodel.tex
\section{Threat Model}
\label{sec:threat_model}

\begin{figure}[t]
    \includegraphics[width=\linewidth]{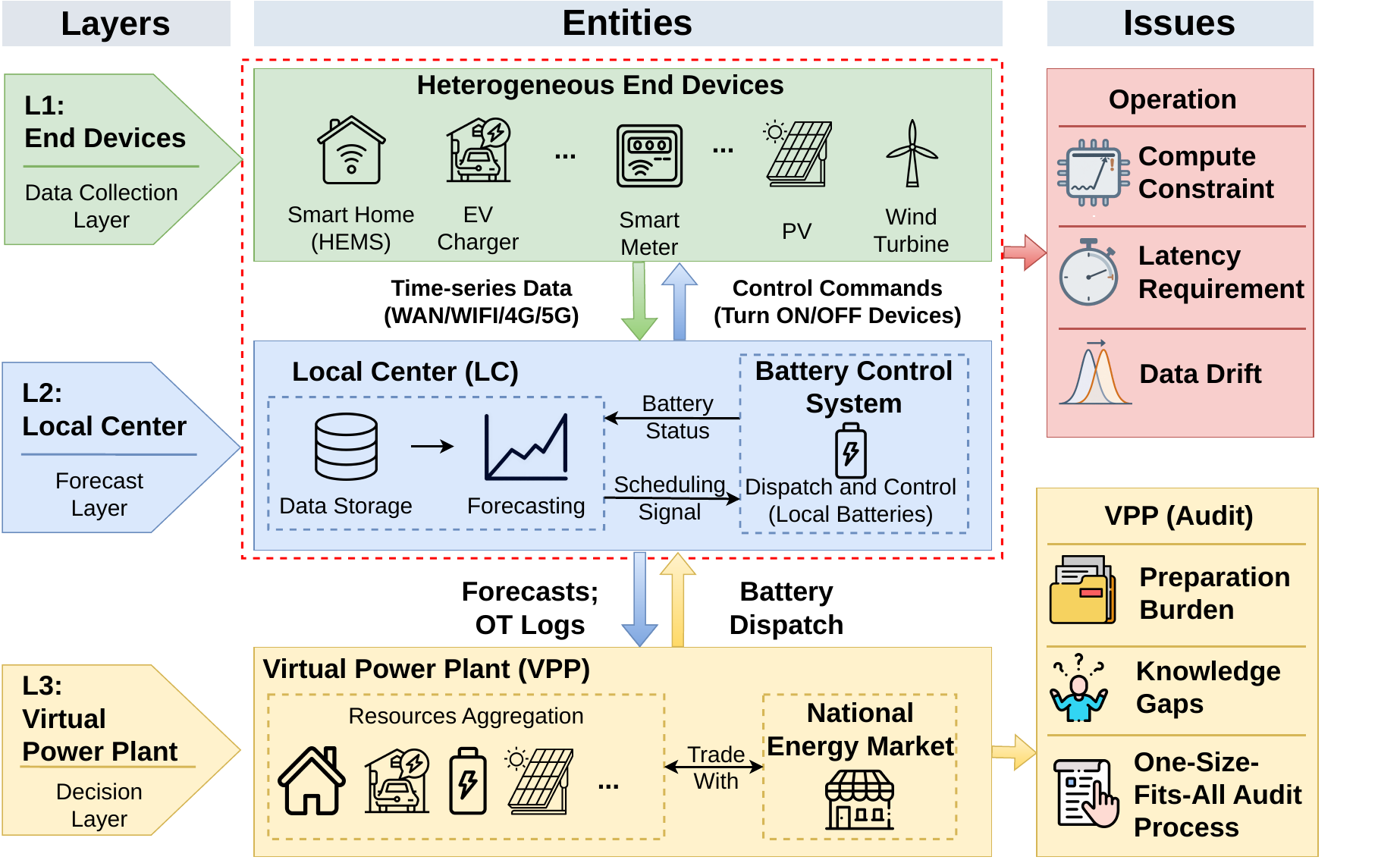}
    \caption{Three-layer energy infrastructure model: entities, cross-layer interactions and issues.}
    \label{fig:framework_overview}
\end{figure}

This section formalizes the three-layer energy infrastructure model and analyzes the entities, data streams, attack surfaces, and attacker objectives at each layer. It then specifies the attacker capabilities relevant to our forecasting defense and log auditing, as well as the operational constraints and security needs of the infrastructure.

\noindent \textbf{Energy infrastructure modeling and attacker's goal.}
\label{subsec:infrastructure_goal}
Energy systems do not prescribe a fixed number of operational tiers, yet a recurring three-layer pattern emerges across major standards and regulatory frameworks: IEC 61850 separates field devices from bay-level and station-level control~\cite{etherden2014vpp}; NIST's smart-grid cybersecurity guidelines stratify assets by communication domain and trust boundary~\cite{nistir7628}; and FERC Order 2222 formalizes the aggregation of distributed energy resources under VPP-like entities that interface with wholesale markets~\cite{ferc2222}. We extract from these frameworks a three-layer conceptual model that captures the hardware, latency, and governance heterogeneity driving distinct security postures. Alternative architectures such as microgrids or community energy systems map onto the same three tiers. Figure~\ref{fig:framework_overview} illustrates the entities, cross-layer interactions, and issues in this model. Telemetry flows from end devices to local centers for forecasting, and the forecasts inform VPP market and dispatch decisions.

Each layer handles distinct data streams, exposes a distinct attack surface, and admits a distinct attacker objective:
\begin{itemize}
    \item \textbf{L1: End devices.} Resource-constrained Advanced Metering Infrastructure (AMI) connected smart meters and IoT sensors reporting raw time-series telemetry. \textit{Attacker's Goal:} Inject perturbed readings to induce downstream forecast deviations large enough to move outside operational tolerance bands.

    \item \textbf{L2: Local centers.} Latency-sensitive control systems (\eg Battery Control System, Energy Management System) that aggregate upstream telemetry and run TS-LLM demand forecasters. \textit{Attacker's Goal:} At this layer, the attacker similarly seeks to distort forecasts through input manipulation, but targets the broader telemetry used for local-center forecasting and operational assessment rather than individual meter readings at L1. To evade detection, the attacker may inject smaller or operationally plausible perturbations, including gradual distribution shifts, to bias downstream market-bidding or battery-dispatch decisions.

    \item \textbf{L3: Virtual power plants (VPPs):} Market-facing entities that aggregate site-level forecasts, Operational Technology (OT) logs, alarms, and balancing orders. \textit{Attacker's Goal:} Flood audit pipelines, forge logs, or execute OT-level command manipulation (\eg Industroyer-class attacks~\cite{app15179233}) to delay attribution and trigger incorrect market actions.
\end{itemize}

\noindent \textbf{Attacker's knowledge and capability.}
\label{subsec:attacker_capability}
\begin{enumerate}[topsep=0pt, partopsep=0pt, itemsep=2pt, parsep=0pt]
    \item \textbf{Black-box access:} At the L1 and L2 layers, the attacker can query the operational TS-LLM forecaster but has no visibility into its parameters, gradients, or clean training data. Adversarial inputs are crafted via gradient-free or generative optimization~\cite{liu2024adversarial, darban2025geniasgeneratorinstantiatinganomalies} (evaluated in \S\ref{sec:evaluation}).

    \item \textbf{Partial compromise:} The attacker controls or impersonates a subset of telemetry or OT communication endpoints (\eg smart meters) but does not obtain administrative control of the forecasting pipeline or the VPP itself.

    \item \textbf{Defense knowledge:} The attacker is aware of the defense mechanism existence and algorithmic logic, but does not know the specific calibrated parameters.
\end{enumerate}

\noindent \textbf{Design requirements.}
\label{subsec:design_requirements}
The defender cannot modify the frozen architecture or weights of the pre-trained TS-LLM, nor does it have access to online ground-truth labels. When scalar adversarial probabilities are required (\eg for the decision-tree mechanism in §\ref{sec:marl}), an independent and lightweight projection layer is appended. 
This head serves as an external defense wrapper rather than a model modification.
The following design requirements reflect the layer-specific constraints and auditing needs summarized in Figure~\ref{fig:framework_overview}:

\begin{itemize}
    \item \textbf{R1 (End devices): Compute constraints.} Resource constrained AMI hardware limits on-device computation. Since all L1 telemetries flow through L2 and the two layers share similar input-manipulation threat, we centralize detection and repair at L2 to address attacks originating at either layer.

    \item \textbf{R2 (Local centers): Latency \& distributional data drift.} The defense must operate online within strict latency budgets (\ie 16ms under IEEE 1646~\cite{ieee1646}) and accommodate benign changes in energy-data distributions over time (\eg changes in renewable generation patterns~\cite{fenza2019drift, lu2019learning}), requiring both detection and immediate signal repair to preserve forecast continuity.
    
    \item \textbf{R3 (VPPs): Scalable forensic auditing.} Large OT log volumes and persistent domain knowledge gaps among auditors~\cite{ccs2025-USElectricPowerSector} require efficient, evidence-grounded assistance for human forensic auditing. The auditing agent should automatically correlate OT logs across devices, identify suspicious events, and organize supporting evidence to reduce preparation effort and help auditors rapidly investigate attacks.
\end{itemize}

%% file: section/4_methodology.tex
\section{Defense Strategies}\label{sec:defense}
This section presents MARL detection and DWT-based repair at local control centers (L2), LLM-assisted forensic auditing at VPPs (L3), and the cross-layer orchestration of our framework.
Building on our model and operational requirements in \S\ref{sec:threat_model}, we first describe how cooperative MARL detection~(\S\ref{sec:marl}) and DWT-based signal repair~(\S\ref{sec:repair}) work together to protect forecast integrity and preserve forecast continuity under local-center latency constraints.
We then present an evidence-grounded LLM forensic auditing agent designed to help human auditors efficiently investigate potential OT attacks (\S\ref{sec:active}).
Finally, we define the data flow and evidence contract connecting end-device telemetry, local-center detection and repair decisions, and downstream VPP operations (\S\ref{sec:cross_layer_contract}).

\subsection{Cooperative Multi-Agent Reinforcement Learning (MARL) Detection}\label{sec:marl}
We adopt a cooperative Multi-Agent Reinforcement Learning (MARL) framework to optimise adversarial screening in the dynamic environment using five category of high-dimensional and heterogeneous features (\eg time, prediction statistics, entropy, metrics, and forecast-model knowledge). These categories provide complementary evidence about temporal context, forecast behavior, prediction consistency, and model responses. Prediction statistics are computed independently on the full and internal predictions defined in the input pre-processing step below, yielding two distinct feature groups and thus six local observations in total (details in Appendix~\ref{apx:feature_vector_construction}). Each observation is assigned to a dedicated, separately parameterized specialist agent, which processes only its assigned feature group and emits an adversarial-confidence message. A coordinating policy fuses the six messages with the decision-tree depth and selects the binary operational action $a_t$, where $a_t=0$ accepts the input and $a_t=1$ rejects it. All six specialists and the coordinator are trained jointly using a shared team reward, while the TS-LLM backbone remains frozen.

\input{tables/alg_decision_tree_mechanism}
\noindent\textbf{Input pre-processing.}
We denote the observed input window as the \emph{full context} and the TS-LLM forecast as the \emph{full prediction}. 
At inference time, the corresponding future observations are unavailable, so the actual forecasting error of the full prediction cannot be used as detection evidence. To obtain prediction-observation comparisons using only available data, we construct an internal forecasting task within the full context. Specifically, we split the full context chronologically into an earlier portion (\ie \emph{internal context}), and a later portion (\ie \emph{internal ground truth}). We front-pad the internal context with zeros to satisfy the forecaster's input-length requirements and feed it into the same frozen TS-LLM to generate the \emph{internal prediction}. Comparing this prediction with the internal ground truth enables the construction of entropy and metric features (details in Appendix~\ref{apx:feature_vector_construction}) without requiring future observations. The resulting six pre-processing streams are: (1) the prediction timestamps, (2) full context, (3) full prediction, (4) internal context, (5) internal prediction, and (6) internal ground truth.

\noindent\textbf{Observations, joint state, and transition.} From the six streams we construct a 100-dimensional feature vector spanning temporal, statistical, and information-theoretic domains (construction details in Appendix~\ref{apx:feature_vector_construction}), fit one anomaly-detection model per feature family whose output is the probability that the input is adversarial under that family's evidence and partition it into the local observations $o_t^1,\ldots,o_t^6$. Specialist agent $i\in\{1,\ldots,6\}$ owns a separately parameterized trainable local policy $h_{\theta_i}$ and emits $p_t^i=h_{\theta_i}(o_t^i)\in[0,1]$. The scalar $p_t^i$ is both the specialist's adversarial confidence and its continuous communication action. The coordinator receives these six messages and concatenates them with the decision-tree depth to form the joint state:
\begin{equation}
s_t=[p_t^1,\ldots,p_t^6,d_t]\in\mathbb{R}^{7},
\qquad d_t\in\{0,1,2\}.
\label{eq:marl-state}
\end{equation}
The coordinator samples the binary operational action $a_t$ from $\pi_{\theta_c}(a_t\mid s_t)$; hence $\mathbf{a}_t=(p_t^1,\ldots,p_t^6,a_t)$ is the hierarchical joint action, comprising six communication actions and one operational action. The depth records how far the decision-tree re-detection procedure escalated before committing to a verdict: $d_t=0$ if the projection-augmented forecast pipeline classified the input as benign on the first pass, $d_t=1$ if a smoothing probe was required to break a suspected perturbation, and $d_t=2$ if an additional odd/even re-split was needed (full procedure in Algorithm~\ref{alg:decision_tree_mechanism} and Appendix~\ref{apx:feature_vector_construction}). A larger $d_t$ therefore simultaneously signals stronger adversarial evidence and a higher compute cost. Windows are processed chronologically. Accepting a window updates a rolling trusted-context buffer with the raw signal; rejecting it updates the buffer only with the repaired signal or the common fallback. Consequently, the joint action changes the context used to construct $s_{t+1}$ rather than reducing the problem to an independently shuffled binary classifier. The detector therefore adjusts its decisions to current evidence and evolving trusted context; this context adaptation does not require online ground-truth labels or imply online policy retraining.

\noindent\textbf{Centralized training via reinforcement learning and label-free deployment.} During offline training, the coordinating MLP policy $\pi_{\theta_c}(a_t\mid s_t)$ samples the binary operational action and the reward $r_t$ scores it against the training-time label $y_t\in\{0,1\}$ (\ie clean vs.\ adversarial) and against the compute already spent, summarised by $d_t$. We define $r_t$ as a function of $(a_t, d_t)$ given $y_t$ in Eq.~\ref{reward_function}, with $a_t$ entering through three outcome-dependent terms and $d_t$ through a single latency term. During deployment, neither the agents nor the coordinator receives $y_t$: they use only $o_t^i$, the joint confidences, and $d_t$. Thus labels define the offline learning signal but are not an online assumption.

\begin{equation}
\label{reward_function}
r_t(a_t, d_t;\, y_t) \;=\; \alpha \cdot \mathrm{AG}(a_t) \;-\; \beta \cdot \mathrm{AM}(a_t, y_t) \;-\; \gamma \cdot \mathrm{FA}(a_t, y_t) \;-\; \delta \cdot \mathrm{LP}(d_t).
\end{equation}

Here \textbf{Accuracy Gain} ($\mathrm{AG}$) rewards the mean-absolute-percentage-error improvement obtained when an adversarial error is avoided, \textbf{Attack Miss} ($\mathrm{AM}$) and \textbf{False Alarm} ($\mathrm{FA}$) penalise accepting adversarial inputs or rejecting clean ones respectively, and the \textbf{Latency Penalty} ($\mathrm{LP}$) charges the compute cost implied by $d_t$ (see Algorithm~\ref{alg:decision_tree_mechanism} in Appendix~\ref{apx:feature_vector_construction}). The policy is therefore incentivised to reject adversarial inputs with minimal overhead while being penalised for both misclassification and for gratuitous escalation on easy samples. All six specialists and the coordinator receive this same scalar team reward. Writing $\Theta=\{\theta_1,\ldots,\theta_6,\theta_c\}$, training optimizes the joint objective
\begin{equation}
J(\Theta)=\mathbb{E}_{\pi_\Theta}\!\left[\sum_{t=0}^{T-1}\rho^t r_t\right]
\label{eq:marl-team-objective}
\end{equation}
where $\rho$ is the discount factor. PPO~\cite{schulman2017proximal-ppo} or GRPO~\cite{shao2024deepseekmath-grpo} updates every parameter block in $\Theta$ jointly; none of the six specialist models is frozen during MARL training (only the TS-LLM backbone remains frozen). This shared-reward, joint-update construction is the cooperative multi-agent component of the detector.

\noindent\textbf{Mitigating distributional data drift in MARL.} The specialists provide complementary evidence that characterizes consistency within the current window. The input features are engineered; joint training learns feature-conditioned confidence mappings and coordination. This combination is intended to help distinguish benign operating changes from perturbations that appear inconsistent across feature families. Diversified offline training exposes the agents to a balanced mix of temporal patterns across datasets, while the false-alarm penalty discourages excessive rejection of legitimate inputs. At deployment, the rolling trusted context changes subsequent observations. These are complementary generalization and context-conditioning mechanisms.

\subsection{Adversarial Input Repair}\label{sec:repair}
Once the MARL detector~(\S\ref{sec:marl}) flags an input as adversarial, the repair stage is invoked and returns a recovered signal on which the downstream TS-LLM can forecast.

\noindent\textbf{Repair goal and metric.} Let $X_{\text{clean}}$, $X_{\text{adv}}$, and $X_{\text{repair}}$ denote the clean, adversarial, and repaired samples. For $Z\in\{X_{\text{adv}},X_{\text{repair}}\}$, we measure distortion relative to the clean sample (Eq~\ref{rse_clean_adv-repair}) using relative squared error (RSE) and the repair objective is to minimize $\mathrm{RSE}(X_{\text{repair}},X_{\text{clean}})$.:

\begin{equation}
\label{rse_clean_adv-repair}
\mathrm{RSE}(Z, X_{\text{clean}})
=
\frac{\sum_{i=1}^{n}\left(X_{\text{clean}}^{(i)}-Z^{(i)}\right)^2}
{\sum_{i=1}^{n}\left(X_{\text{clean}}^{(i)}-\bar{X}_{\text{clean}}\right)^2}.
\end{equation}

\noindent\textbf{Discrete wavelet transform~(DWT) and parameter scope.} We use a fixed four-level ($J=4$) Discrete Wavelet Transform (DWT) across all samples. The transform decomposes a signal into one low-frequency approximation band and $J$ high-frequency detail bands; soft-thresholding the detail coefficients and reconstructing the signal suppresses high-frequency perturbations while retaining the dominant consumption trend. We write $\mathrm{Repair}(X_{\mathrm{adv}},P)\to X_{\mathrm{repair}}$, where $P=(\psi_d,\{\lambda_j\}_{j=1}^{J},\alpha_u)$. The wavelet family $\psi_d$ is selected once from the clean training partition for dataset $d$, and the blending ratio $\alpha_u$ is calibrated only from user $u$'s clean training history. At runtime, these quantities and $J$ remain fixed; for each rejected window, the repair stage retrieves a user-local clean reference and adapts only the per-level soft-threshold multipliers $\{\lambda_j\}$.

\noindent\textbf{User-local reference retrieval and per-window threshold calibration.} Under the black-box threat model of \S\ref{sec:threat_model}, the repair stage cannot query, simulate, or synthesize attack-specific samples; its calibration anchor must therefore be derived entirely from clean historical data already held by the operator. At onboarding, for each user $u$, the operator constructs a user-local historical pool $X_{\mathrm{pool}}^{(u)}$ exclusively from that user's clean training-period data and caps the pool at 10,000 sliding windows. Each user-local pool is indexed offline using a Hierarchical Navigable Small World (HNSW) graph~\cite{malkov2018efficient} for approximate nearest-neighbor retrieval. When the MARL coordinator rejects an incoming window $X_{\mathrm{adv}}^{(u)}$, the HNSW search returns an approximate RSE-nearest clean reference:
\begin{equation}
\widehat{i}
=
\operatorname{HNSW}_{\mathrm{RSE}}\!\left(
X_{\mathrm{adv}}^{(u)};
X_{\mathrm{pool}}^{(u)}
\right),
\qquad
\widehat{X}_{\mathrm{clean-ref}}
=
X_{\mathrm{pool},\widehat{i}}^{(u)}\ .
\end{equation}
Here, $\operatorname{HNSW}_{\mathrm{RSE}}$ denotes an approximate search for the reference minimizing $\mathrm{RSE}(X_{\mathrm{adv}}^{(u)},X_{\mathrm{pool},i}^{(u)})$. With the decomposition depth $J$, dataset-level wavelet family $\psi_d$, and user-level blending ratio $\alpha_u$ fixed as described above, the repair stage adapts only the per-level soft-threshold multipliers for the current rejected window:
\begin{equation}
\{\lambda_j^\star\}_{j=1}^{J}
=
\arg\min_{\{\lambda_j\}_{j=1}^{J}}
\mathrm{RSE}\!\left(
\mathrm{Repair}\!\left(
X_{\mathrm{adv}}^{(u)},
(\psi_d,\{\lambda_j\}_{j=1}^{J},\alpha_u)
\right),
\widehat{X}_{\mathrm{clean-ref}}
\right).
\end{equation}
This per-window calibration allows the shrinkage at each decomposition level to respond to the scale-specific differences between the rejected window and its retrieved clean reference. The resulting signal,
\begin{equation}
X_{\mathrm{repair}}
=
\mathrm{Repair}\!\left(
X_{\mathrm{adv}}^{(u)},
(\psi_d,\{\lambda_j^\star\}_{j=1}^{J},\alpha_u)
\right),
\end{equation}

is forwarded to the downstream forecaster in place of $X_{\mathrm{adv}}^{(u)}$. Because the procedure uses only the user's clean training history, the current rejected window, and its retrieved reference, it requires neither the attacker's algorithm or attack identity, nor a test label or the unavailable clean counterpart of the attacked window; it is therefore attack-agnostic under our black-box threat model. Crucially, $\widehat{X}_{\mathrm{clean-ref}}$ is used strictly as a calibration anchor and is never substituted for the live input.

\noindent\textbf{Role in mitigating distributional data drift.} Selecting a reference for the current window and recalibrating its thresholds avoids using one fixed shrinkage setting for all operating patterns. The historical reference may not fully capture the current operating context (\eg weather-driven load, scheduling shifts, or behavioral variations) that remains present in $X_{\mathrm{adv}}^{(u)}$; forwarding it directly would replay an outdated pattern. Repair therefore remains centered on the live input, with reference-calibrated detail-band shrinkage and blending intended to retain benign temporal structure while suppressing perturbations. The reference pool remains fixed after on-boarding: this is per-window parameter calibration, not online acquisition of new clean samples. Its ability to accommodate drift depends on representative same-user history; an unseen regime may lack a suitable reference, motivating the confidence gate below. The reported DWT-latency includes HNSW retrieval, per-level threshold calibration, inverse-DWT reconstruction, and blending.

\noindent\textbf{Reference-confidence gate.} Before per-window threshold calibration, we permit repair only when a reference is available and $\mathrm{RSE}(X_{\mathrm{adv}}^{(u)},\widehat{X}_{\mathrm{clean-ref}})\leq\tau_u$. The user-specific threshold $\tau_u$ is a preselected upper quantile of nearest-reference distances from clean historical calibration windows, disjoint and non-overlapping with the reference pool. Calibration uses the same retrieval and distance computation as deployment, with a fixed positive denominator floor for near-constant reference windows; no test labels or attack identities are used. If the gate fails or history is unavailable, we abstain from DWT repair and use the common persistence forecast. This gate screens reference mismatch rather than guaranteeing repair safety; the un-gated variant omits the distance check.

\subsection{LLM Forensic Auditing Agent}
\label{sec:active}

To accelerate forensic auditing and support accurate investigation of potential OT attacks at VPPs, we deploy an LLM-based forensic auditing agent that operates over device error logs from the target infrastructure. The agent's workflow consists of four stages:

\begin{itemize}
    \item \textbf{Input ingestion.}
    The agent receives an accumulated OT-log window $\mathcal{L}$, together with a run-time candidate-attack set $\mathcal{A}$ describing the attack types relevant to the deployment.
    In the Sherlock evaluation, this set contains Industroyer, Drift-Off, Control \& Freeze, and ARP Spoofing DoS.

    \item \textbf{Reasoning and correlation.}
    The agent analyzes log entries across devices and time windows, correlating anomalous patterns (e.g., repeated OT command sequences, measurement discontinuities, and communication disruptions) to reconstruct candidate attack narratives.
    Attack hypotheses are restricted to types in $\mathcal{A}$.

    \item \textbf{Evidence-grounded reporting.}
    To mitigate hallucinations and unsupported claims, every anomaly flag must be grounded in specific log entries, including timestamps, device identifiers, and log content.
    Each proposed attack hypothesis must be supported by this cited evidence.

    \item \textbf{Escalation output.}
    The agent produces a structured report containing the supported attack hypothesis, log evidence, affected assets, and a recommended action. It returns an empty anomaly list if the logs support no candidate type.

\end{itemize}

The agent operates as a forensic filter within a human-in-the-loop (HITL) system: it surfaces potential threats with supporting evidence and recommends escalation, but cannot autonomously execute remediation actions. Final validation and control-action authority remain with human auditors. This design is intended to reduce the evidence-preparation burden and help bridge the OT-domain knowledge gap identified by Sahin et al.

\subsection{Cross-Layer Orchestration and Evidence Contract}\label{sec:cross_layer_contract} The components described above are connected through a cross-layer data flow (as shown in Figure~\ref{fig:framework_overview}) and a common evidence contract that records their processing decisions and downstream consequences. End devices (L1) forward telemetry to local centers (L2), where the MARL detector screens each input window. Accepted windows are used directly for forecasting, whereas rejected windows undergo DWT-based repair when permitted by the reference-confidence gate. If the gate abstains, the system uses a persistence forecast to preserve forecast continuity; the same fallback is used across all ablations. The resulting forecast is then forwarded to the VPP (L3), where it informs the net-load bid and subsequent operational decisions.

To make this process traceable, each input window is assigned an event identifier and a hash of the received values. At L2, the record is extended with the six specialist confidences, local actions, coordinator action, and decision-tree depth. Repair-related fields include the selected clean-reference identifier and distance, DWT parameters, repaired-signal hash, and before/after forecasts, where applicable. At L3, the VPP appends the forecast-derived net-load bid and any operator action. These records are serialized as timestamped entries in the VPP audit log.

All entries associated with the same telemetry window retain the same event identifier. This linkage allows operators to trace an L1 observation through its L2 detection and repair decisions to its recorded L3 consequences, without requiring access to complete raw user histories for record association. The evidence contract therefore specifies how cross-layer records are organized and linked, while the forensic-auditing workflow in \S\ref{sec:active} describes how OT logs are analyzed to support human investigation and validation.

%% file: tables/alg_decision_tree_mechanism.tex
\begin{algorithm}[t]
\caption{Decision tree mechanism for adversarial detection}
\label{alg:decision_tree_mechanism}
\resizebox{\linewidth}{!}{%
\begin{minipage}{1.06\linewidth}
\begin{algorithmic}[1]
\Require Full context $X$; forecasting pipeline $f$ (frozen backbone with appended projection head); DWT-based smoother $\mathcal{S}$
\Ensure Detection probability $\mathrm{Prob}$; depth $d$

\State $\mathrm{Prob}_0 \gets f(X)$
\State \textbf{if} $\mathrm{Prob}_0 < 0.5$ \textbf{then} \Return $(\mathrm{Prob}_0, 0)$
\State $\tilde{X} \gets \mathcal{S}(X)$
\State \textbf{if} $f(\tilde{X}) < 0.5$ \textbf{then} \Return $(\mathrm{Prob}_0, 1)$
\State Split $\tilde{X}$ into odd/even subsequences $X^{\mathrm{odd}}, X^{\mathrm{even}}$
\State Front-pad both subsequences with zeros to length $|X|$
\State $\mathrm{Prob}_2^{\mathrm{odd}} \gets f(X^{\mathrm{odd}})$,
       $\mathrm{Prob}_2^{\mathrm{even}} \gets f(X^{\mathrm{even}})$
\If{$\max(\mathrm{Prob}_2^{\mathrm{odd}}, \mathrm{Prob}_2^{\mathrm{even}}) < 0.5$}
    \State \Return $(\mathrm{Prob}_2^{\mathrm{odd}} + \mathrm{Prob}_2^{\mathrm{even}}, 2)$
\EndIf
\State \Return $(\mathrm{Prob}_0, 2)$

\end{algorithmic}
\end{minipage}
}
\end{algorithm}

%% file: section/5_experiment_and_evaluation.tex
\section{Experiment and Evaluation}\label{sec:evaluation}
The full experiment setup is documented in Appendix~\ref{apx:exp_setup}. In this section, we first evaluate MARL detection performance (\S\ref{sec:eval_detection}), then compare coordination strategies (\S\ref{sec:eval_fusion_baselines}), assess input repair and downstream forecast recovery (\S\ref{sec:eval_repair}), and examine DWT repair baselines and robustness to reference-history conditions (\S\ref{sec:as_dww_repair_and_robustness}). We separately evaluate VPP log forensic auditing on the Sherlock~\cite{Sherlock} benchmark (\S\ref{sec:active_eval}) and finally measure L2 runtime overhead (\S\ref{sec:eval_perf}).

\subsection{Adversarial Detection Performance}\label{sec:eval_detection}
\input{tables/adversarial_detection}
We evaluate the MARL detector (\S\ref{sec:marl}) using PPO~\cite{schulman2017proximal-ppo} and GRPO~\cite{shao2024deepseekmath-grpo}, both with a learning rate of $10^{-3}$. During training, each RL-algorithm jointly updates all six specialist models and the coordinator; at inference time, none of them receives a ground-truth label. We use TimesFM~\cite{das2024decoder-timesfm} and TimeLLM~\cite{jin2023timellm-etth2} as frozen forecasting backbones and train the MARL detector around each backbone without updating its parameters. In contrast, TimeGPT~\cite{garza2023timegpt} is a closed-source commercial forecaster with a native anomaly-detection endpoint. We evaluate this endpoint directly on the same held-out windows as an independent baseline.

Table~\ref{tab:adversarial_detection_performance} reports the F1 scores on the held-out test sets. Our best configuration, GRPO+TimeLLM, achieves F1\,$>$\,0.91 on every dataset--attack pair. Under DGA, it reaches 0.9923, 0.9911, and 0.9198 on NI, HECP, and Electricity, respectively; the corresponding scores under GenIAS~\cite{darban2025geniasgeneratorinstantiatinganomalies} are 0.9996, 1.0000, and 0.9841. TimesFM-based configurations achieve F1 scores of 0.7778--0.7882 on HECP~\cite{hecp} and Electricity~\cite{trindade2015electricity} under DGA. However, TimeLLM's advantage is not consistent across optimizers: on HECP under DGA, PPO+TimeLLM scores 0.7778, slightly below PPO+TimesFM (0.7867). All four MARL configurations outperform the TimeGPT~\cite{garza2023timegpt} commercial native anomaly-detection baseline on every dataset--attack pair, whose F1 never exceeds 0.3156.

\begin{takeaway}
\textbf{Takeaway 1:~}
\textit{Our MARL detector, built on open-source TS-LLMs wrapped in a RL policy over the \S\ref{sec:marl} feature vector, outperforms the commercial baseline on all three datasets under both DGA and GenIAS attack; the best configuration (GRPO~+~TimeLLM) achieves F1\,$>$\,0.91 across every dataset--attack pair.}
\end{takeaway}

\subsection{Coordination Strategies Comparison}\label{sec:eval_fusion_baselines}
\input{tables/fusion_baseline}
To compare coordination strategies, every method uses the same six local observation definitions, decision-tree-depth input, and evaluation splits. Trainable components are optimized under each method's own objective. Because the full MARL configurations jointly update the six specialists and the coordinator, their post-training confidence messages are not constrained to be numerically identical to those of the non-RL baselines. Splits are chronological for region datasets and user-held-out for Electricity~\cite{trindade2015electricity}, preventing overlapping-window leakage.

Table~\ref{tab:fusion_baselines} therefore reports a matched-input end-to-end coordination comparison rather than a frozen same-state fusion-only ablation. The cost combines normalized penalties for attack misses, false alarms, and latency.GRPO achieves the highest F1 (0.965), reduces FPR from 0.034 for the supervised MLP to 0.022, and lowers normalized operational cost from 0.080 to 0.059.

\subsection{Repair Performance}\label{sec:eval_repair}
\input{tables/repair_forecast_compare}

We select the data-level wavelet family from the clean training samples and evaluate the DWT repair procedure (\S\ref{sec:repair}) under DGA~\cite{liu2024adversarial} and GenIAS~\cite{darban2025geniasgeneratorinstantiatinganomalies}, assessing both input recovery and downstream forecast recovery. 

The \textit{repair/*} rows of Table~\ref{tab:dwt_repair_performance} compare adversarial and repaired inputs against their clean counterparts. DWT repair reduces input-space RSE in all six dataset--attack settings, by 34.73\%--60.74\% under DGA~\cite{liu2024adversarial} and 38.91\%--83.36\% under GenIAS~\cite{darban2025geniasgeneratorinstantiatinganomalies}. We next assess downstream recovery by comparing forecasts from adversarial inputs ($P_{adv}$) and repaired inputs ($P_{smooth}$) against the clean-input forecast ($P_{clean}$). Forecast-space RSE decreases in every evaluated setting. TimeLLM~\cite{jin2023timellm-etth2} achieves reductions of 39.04\%--63.04\% under DGA and 39.91\%--89.04\% under GenIAS~\cite{darban2025geniasgeneratorinstantiatinganomalies}; the corresponding ranges for TimesFM~\cite{das2024decoder-timesfm} are 9.39\%--44.75\% and 9.70\%--22.09\%. TimeLLM~\cite{jin2023timellm-etth2} shows larger relative reductions in all six settings.

\begin{takeaway}
\textbf{Takeaway 2:~}
\textit{DWT repair reduces input-space RSE by 34.73\%--83.36\% and lowers forecast-space RSE by up to 89.04\% on TimeLLM and 44.75\% on TimesFM, relative to their respective clean baselines.}
\end{takeaway}

\subsection{Repair Baselines and Robustness}\label{sec:as_dww_repair_and_robustness}
\input{tables/dwt_baseline}
\input{tables/dwt_stress}

\noindent\textbf{Evaluation protocol.} We evaluate each repair method on actual detector rejections, including clean false positives when measuring clean distortion and harm. Test labels, attack identities, and clean counterparts of rejected windows are used only for post-hoc evaluation. The primary DWT setting uses 10K same-user clean training windows under in-season conditions, with reference construction and calibration following \S\ref{sec:repair}. Harm is the fraction of repaired samples whose downstream forecast error increases by more than 1\% after repair.

\noindent\textbf{Repair baselines and confidence gating.} Table~\ref{tab:dwt_baselines} compares the adaptive DWT design with two generic signal smoothers and a fixed-parameter DWT. Savitzky--Golay filtering and Kalman smoothing have the lowest latency (0.31\,ms and 0.44\,ms), but provide only 31.6\% and 35.8\% RSE reduction, respectively, while producing clean distortion of 2.9\% and 2.4\% and harm rates of 12.7\% and 10.8\%. Fixed DWT improves the RSE reduction to 42.6\% and lowers clean distortion and harm to 2.0\% and 8.1\%, respectively, at 0.61\,ms. Adapting the four per-level thresholds increases the RSE reduction to 63.4\% without gating, while reducing clean distortion to 1.3\% and harm to 5.4\%. Adding the confidence gate sacrifices only 1.6 percentage points of RSE reduction (63.4\% to 61.8\%) and adds 0.03\,ms of latency (2.32\,ms to 2.35\,ms), but reduces clean distortion from 1.3\% to 0.6\% and harm from 5.4\% to 2.1\%. Thus, per-window threshold adaptation provides the principal repair gain, while confidence gating substantially limits repair-induced damage on unreliable references and false-positive detector outputs.

\noindent\textbf{Reference-pool robustness.} Table~\ref{tab:dwt_stress} examines sensitivity to reference-pool size, history relation, and operating condition. For gated DWT, repair coverage is the fraction of detector rejections passing the reference-confidence gate. Under same-user, in-season matching, increasing the historical pool from 100 to 1K and 10K windows raises RSE reduction from 50.9\% to 58.7\% and 61.8\%, while repair coverage increases from 86.0\% to 91.7\% and 94.2\%. Over the same progression, clean distortion decreases from 1.1\% to 0.8\% and 0.6\%, and harm falls from 4.6\% to 3.0\% and 2.1\%. This improvement incurs a moderate latency increase from 1.71\,ms to 1.98\,ms and 2.35\,ms. Holding the pool size at 10K isolates reference mismatch: cross-user matching reduces RSE reduction to 45.6\%, increases harm to 6.8\%, and lowers coverage to 78.4\%; seasonal shift further reduces RSE reduction to 42.8\%, with 8.4\% harm and 72.6\% coverage; and rare-event windows are the most difficult non-missing-history condition, yielding 29.4\% RSE reduction, 9.7\% harm, and 55.1\% coverage. Their nearly identical latencies (2.34--2.36\,ms) show that these degradations arise from reference mismatch rather than additional computation. When no historical reference is available, the gate performs no DWT repair and invokes the common fallback, producing 0.0\% repair coverage, RSE reduction, clean distortion, and repair-induced harm at 0.08\,ms. These results expose the method's dependence on representative same-user history while showing that the confidence gate reduces coverage instead of forcing repair with an unreliable reference.

\subsection{Audit Agent Performance}\label{sec:active_eval}
\input{tables/attack_detection_from_device_log}
We evaluate the LLM auditing agent on the Sherlock benchmark~\cite{Sherlock}, which was generated with the Wattson simulator~\cite{bader2023comprehensively} and contains over 30 days of power-grid behavior in basic, semi-urban, and rural scenarios. We use the basic test set to evaluate both zero-shot and few-shot learning and the semi-urban test set to provide few-shot examples, as the training data contain no attacks. We examine attack-type coverage, the detection--false-alert trade-off, and the final configuration's sensitivity to degraded logs and prompt-example ordering.

The Sherlock~\cite{Sherlock} benchmark covers four OT attack types: \textit{Industroyer} disrupts grid operations through malicious commands; \textit{Drift-Off} uses a man-in-the-middle (MITM) attack to manipulate measurements and mislead the control center (CC); \textit{Control \& Freeze} conceals grid disruptions through MITM replay and interpolation of recorded measurement trends; and \textit{ARP Spoofing DoS} disrupts CC--remote terminal unit communications through ARP spoofing.

The agent we employ is Claude Code, an agentic LLM assistant with strong code and log comprehension capabilities. For zero-shot evaluation, we provide the four-type runtime candidate set and the device error log; for few-shot learning, prompts include additional examples drawn from the semi-urban test set (log files and ground-truth attack labels). Consistent with the human-in-the-loop design of \S\ref{sec:active} (\ie the agent surfaces, the human authorizes), both settings ran the agent in plan mode: the agent drafted an analysis plan and waited for operator approval before executing it. In the evidence-gated few-shot configuration, a proposed anomaly flag is retained for reporting only if its cited timestamp, device identifier, and log line match the input log, following the evidence-grounding constraint of \S\ref{sec:active}.

\input{tables/audit_eval}

Table~\ref{tab:attack_detection_devices_log} presents detections by attack type on 18 attack windows. The agent detects 9/18 in the zero-shot setting (50\%) and 14/18 with few-shot examples (77.8\%). The largest gain is for Control \& Freeze (1/5 to 4/5); Industroyer improves from 6/7 to 7/7, whereas Drift-Off remains at 2/4 and ARP Spoofing DoS reaches only 1/2. Thus, the added examples improve attack coverage unevenly across families.

Attack-only detection rates do not capture the false alerts that human auditors must review. Table~\ref{tab:audit_eval} therefore includes 180 benign windows alongside the 18 attack windows and compares rule-based signatures, a single-pass LLM, plan-mode zero-shot, and the few-shot evidence-gated agent. The final configuration detects 14 attacks with 5 false positives (73.7\% precision, 77.8\% recall, 2.8\% FPR, and 75.7\% F1), compared with 13 true positives and 16 false positives for the single-pass LLM, and 11 and 8, respectively, for rule-based signatures. Plan-mode zero-shot alone detects only 9 attacks with 12 false positives; the observed advantage is for the complete few-shot evidence-gated configuration, not for planning in isolation. Its 5 false alerts over the 30 evaluated days (0.17/day) remain a direct review-burden proxy rather than a negligible cost.

\input{tables/audit_robustness}
Performance on intact logs does not establish behavior when audit evidence is incomplete or corrupted. Table~\ref{tab:audit_robustness} fixes the few-shot evidence-gated configuration and tests it over three runs with missing lines, token noise, truncated logs, or shuffled prompt examples (not reordered log lines). With 25\% truncation, recall falls from $77.8{\pm}3.2$\% to $66.7{\pm}3.2$\% and abstention rises from 4.8\% to 17.4\%; with 10\% token noise, FPR rises from $2.8{\pm}0.6$\% to $3.5{\pm}0.9$\%. Evidence-valid is the fraction of retained flags that pass the strict timestamp/device/log-line parser after filtering. The reported 100\% is conditional on retained flags; it does not measure validity among raw proposed flags or correctness of the attack diagnosis. Unmatched citations are withheld from escalation, and retained reports remain subject to human validation before any control action.

\begin{takeaway}
\textbf{Takeaway 3:~}
\textit{Our forensic auditing pairs four-stage LLM auditing agent with human-in-the-loop validation, catching 77.78\% of OT-level attacks overall and 100\% of Industroyer-class incidents in the few-shot setting, directly addressing the preparation burden and OT-domain knowledge gap.}
\end{takeaway}

\subsection{Performance Overhead}\label{sec:eval_perf}
The full GRPO+TimeLLM MARL+DWT detection-and-repair pipeline, running under our workstation (AMD Ryzen~9 9950X+NVIDIA RTX~6000 Ada), completes in 7.95\,ms (mean) and 14.50\,ms (worst observed) on every sampled window across all three datasets, strictly below the 16\,ms budget. 

%% file: tables/adversarial_detection.tex
\begin{table}[t]
\centering
\caption{The performance of adversarial detection.}
\label{tab:adversarial_detection_performance}
\resizebox{\linewidth}{!}{
    \begin{tabular}{llccccc}
    \toprule
    \multirow{2}{*}{\textbf{\makecell[l]{Attack\\Approach}}} &
    \multirow{2}{*}{\textbf{Dataset (\# Samples)}} &
    \multicolumn{2}{c}{\textbf{TimesFM~\cite{das2024decoder-timesfm}}} &
    \multicolumn{2}{c}{\textbf{TimeLLM~\cite{jin2023timellm-etth2}}} &
    \multirow{2}{*}{\textbf{\makecell{TimeGPT~\cite{garza2023timegpt}\\(Native AD)}}} \\
    \cmidrule(lr){3-4} \cmidrule(lr){5-6}
    & & \textbf{PPO~\cite{schulman2017proximal-ppo}} & \textbf{GRPO~\cite{shao2024deepseekmath-grpo}} & \textbf{PPO~\cite{schulman2017proximal-ppo}} & \textbf{GRPO~\cite{shao2024deepseekmath-grpo}} & \\
    \midrule
    \multirow{3}{*}{DGA~\cite{liu2024adversarial}}
    & NI~\cite{nidataset} (9058) & 0.9124 & \textbf{\textcolor{red}{0.9375}} & 0.9144 & \textbf{\textcolor{red}{\underline{0.9923}}} & 0.1536 \\
    & HECP~\cite{hecp} (450) & \textbf{\textcolor{red}{0.7867}} & 0.7778 & 0.7778 & \textbf{\textcolor{red}{\underline{0.9911}}} & 0.3156 \\
    & Electricity~\cite{trindade2015electricity} (8672) & 0.7866 & \textcolor{red}{\textbf{0.7882}} & 0.8721 & \textbf{\textcolor{red}{\underline{0.9198}}} & 0.2312 \\

    \midrule
    \multirow{3}{*}{GenIAS~\cite{darban2025geniasgeneratorinstantiatinganomalies}}
    & NI~\cite{nidataset} (9058) & 0.9859 & \textbf{\textcolor{red}{0.9918}} & 0.9898 & \textbf{\textcolor{red}{\underline{0.9996}}} & 0.2527 \\
    & HECP~\cite{hecp} (450) & 0.9778 & \textbf{\textcolor{red}{0.9822}} & 0.9778 & \textbf{\textcolor{red}{\underline{1.0000}}} & 0.1622 \\
    & Electricity~\cite{trindade2015electricity} (8672) & 0.9137 & \textcolor{red}{\textbf{0.9458}} & 0.9195 & \textbf{\textcolor{red}{\underline{0.9841}}} & 0.3149 \\
    \bottomrule \\
    \end{tabular}
}

Note: AD refers to anomaly-detection.
\end{table}

%% file: tables/fusion_baseline.tex
\begin{table}[t]
\centering
\caption{Coordination strategy comparison}
\label{tab:fusion_baselines}
\resizebox{\linewidth}{!}{
\begin{tabular}{lccccccr}
\toprule
\textbf{Fusion method} & \textbf{\makecell{Precision \\($\uparrow$)}} & \textbf{\makecell{Recall\\($\uparrow$)}} & \textbf{\makecell{FPR\\($\downarrow$)}} & \textbf{\makecell{F1\\($\uparrow$)}} & \textbf{\makecell{AUPRC\\($\uparrow$)}} & \textbf{\makecell{Latency\\(ms)\\($\downarrow$)}} & \textbf{\makecell{Cost\\($\downarrow$)}} \\
\midrule
Majority Vote & 0.889 & 0.861 & 0.108 & 0.875 & 0.901 & 0.006 & 0.168 \\
Weighted Vote & 0.913 & 0.899 & 0.076 & 0.906 & 0.929 & 0.009 & 0.133 \\
Logistic Regression & 0.925 & 0.916 & 0.062 & 0.920 & 0.945 & 0.025 & 0.116 \\
Random Forest & 0.942 & 0.934 & 0.045 & 0.938 & 0.963 & 0.188 & 0.094 \\
XGBoost & 0.950 & 0.943 & 0.038 & 0.946 & 0.972 & 0.154 & 0.086 \\
Supervised MLP & 0.954 & 0.948 & 0.034 & 0.951 & 0.977 & 0.066 & 0.080 \\
MARL (PPO) & 0.960 & 0.954 & 0.028 & 0.957 & 0.982 & 0.082 & 0.070 \\
MARL (GRPO) & \textbf{0.969} & \textbf{0.962} & \textbf{0.022} & \textbf{0.965} & \textbf{0.987} & 0.087 & \textbf{0.059} \\
\bottomrule
\end{tabular}}
\end{table}

%% file: tables/repair_forecast_compare.tex
\begin{table*}[t]
\centering
\caption{DWT repair performance: input-space RSE (\textit{repair/*} rows) and forecast-space RSE (TimesFM~\cite{das2024decoder-timesfm}/TimeLLM~\cite{jin2023timellm-etth2} rows), each measured relative to its clean baseline.}
\label{tab:dwt_repair_performance}
\begingroup
\renewcommand{\arraystretch}{0.95}
\setlength{\tabcolsep}{4pt}
\resizebox{0.9\linewidth}{!}{
\begin{tabular}{cccccccccc}
\toprule
\multirow{2}{*}{Dataset (\# Samples)} &
\multirow{2}{*}{Component} &
\multicolumn{4}{c}{DGA~\cite{liu2024adversarial}} &
\multicolumn{4}{c}{GenIAS~\cite{darban2025geniasgeneratorinstantiatinganomalies}} \\
\cmidrule(lr){3-6} \cmidrule(lr){7-10}
& &
Adv. RSE & Smoothed RSE & Gap & Reduction (\%) &
Adv. RSE & Smoothed RSE & Gap & Reduction (\%) \\
\midrule

\multirow{3}{*}{NI~\cite{nidataset} (22645)}
& \textit{repair (sym2)}
& 2.4239 & 1.5797 & -0.8441 & 34.8267
& 11.5448 & 1.9207 & -9.6241 & 83.3629 \\
& TimesFM
& 0.6583 & 0.5947 & -0.0636 & 9.6600
& 1.2777 & 0.9955 & -0.2822 & 22.0900 \\
& TimeLLM
& 0.8323 & 0.4168 & -0.4156 & 49.9300
& 10.4169 & 1.1412 & -9.2757 & 89.0400 \\
\midrule

\multirow{3}{*}{HECP~\cite{hecp} (1123)}
& \textit{repair (bior2.4)}
& 2.5202 & 1.6449 & -0.8753 & 34.7301
& 9.5244 & 1.7693 & -7.7551 & 81.4237 \\
& TimesFM
& 0.5175 & 0.4690 & -0.0486 & 9.3880
& 1.1124 & 0.9617 & -0.1507 & 13.5500 \\
& TimeLLM
& 1.1861 & 0.7230 & -0.4631 & 39.0400
& 5.4159 & 0.6298 & -4.7861 & 88.3700 \\
\midrule

\multirow{3}{*}{Electricity~\cite{trindade2015electricity} (21680)}
& \textit{repair (coif1)}
& 519362.7782 & 203883.7931 & -315478.9851 & 60.7435
& 144.0425 & 87.9886 & -56.0540 & 38.9149 \\
& TimesFM
& 3626.8160 & 2003.9780 & -1622.8380 & 44.75
& 452258.1052 & 408389.6982 & -43868.4070 & 9.7000 \\
& TimeLLM
& 25.4065 & 9.3905 & -16.0160 & 63.0400
& 196.0906 & 117.8389 & -78.2517 & 39.9100 \\

\bottomrule
\end{tabular}
}
\endgroup
\end{table*}

%% file: tables/dwt_baseline.tex
\begin{table}[t]
\centering
\caption{Repair baselines on actual detector outputs.}
\label{tab:dwt_baselines}
\resizebox{0.85\linewidth}{!}{
\begin{tabular}{lcccc}
\toprule
\textbf{Repair} & \textbf{\makecell{RSE\\Reduction \\$\uparrow$}} & \textbf{\makecell{Clean\\Distortion\\$\downarrow$}} & \textbf{Harm $\downarrow$} & \textbf{\makecell{Latency \\ (ms)}} \\
\midrule
Savitzky--Golay & 31.6\% & 2.9\% & 12.7\% & 0.31 \\
Kalman Smoother & 35.8\% & 2.4\% & 10.8\% & 0.44 \\
Fixed DWT & 42.6\% & 2.0\% & 8.1\% & 0.61 \\
Adaptive DWT (w/o gate) & \textbf{63.4\%} & 1.3\% & 5.4\% & 2.32 \\
Adaptive DWT (w gate) & 61.8\% & \textbf{0.6\%} & \textbf{2.1\%} & 2.35 \\
\bottomrule
\end{tabular}}
\end{table}

%% file: tables/dwt_stress.tex
\begin{table}[t]
\centering
\caption{Adaptive-DWT confidence-gate stress test across reference-pool sizes, history relations, and operating conditions. Cross-user matching is a stress-test condition only.}
\label{tab:dwt_stress}
\resizebox{\linewidth}{!}{
\begin{tabular}{cccccccc}
\toprule
\textbf{Pool} & \textbf{\makecell{History\\Relation}} & \textbf{\makecell{Operating\\Condition}} & \textbf{\makecell{RSE \\Reduction\\ ($\uparrow$)}} & \textbf{\makecell{Clean \\Distortion\\ ($\downarrow$)}} & \textbf{\makecell{Harm\\($\downarrow$)}} & \textbf{\makecell{Repair\\Coverage}} & \textbf{\makecell{Latency\\(ms)}} \\
\midrule
100 & Same user & In-season & 50.9\% & 1.1\% & 4.6\% & 86.0\% & 1.71 \\
1K & Same user & In-season & 58.7\% & 0.8\% & 3.0\% & 91.7\% & 1.98 \\
10K & Same user & In-season & \textbf{61.8\%} & \textbf{0.6\%} & \textbf{2.1\%} & 94.2\% & 2.35 \\
10K & Cross user & In-season & 45.6\% & 1.7\% & 6.8\% & 78.4\% & 2.34 \\
10K & Same user & Seasonal shift & 42.8\% & 2.1\% & 8.4\% & 72.6\% & 2.36 \\
10K & Same user & Rare-event window & 29.4\% & 2.6\% & 9.7\% & 55.1\% & 2.35 \\
0 & Missing history & Any & 0.0\% & 0.0\% & 0.0\% & 0.0\% & 0.08 \\
\bottomrule
\end{tabular}}
\end{table}

%% file: tables/attack_detection_from_device_log.tex
\begin{table}[t]
\centering
\caption{Performance on attack detection from devices log}
\resizebox{0.75\linewidth}{!}{
\begin{tabular}{lccc}
\toprule
\textbf{Attack Type} & \textbf{\# Samples} & \textbf{Zero Shot} & \textbf{Few Shot} \\
\midrule
Industroyer&7&6&7\\
Drift Off&4&2&2\\
Control \& Freeze&5&1&4\\
ARP Spoofing DOS&2&0&1\\
\midrule
Total&18&9&14\\
\bottomrule
\end{tabular}
}
\label{tab:attack_detection_devices_log}
\end{table}

%% file: tables/audit_eval.tex
\begin{table}[t]
\centering
\caption{Audit evaluation on 18 attack windows and 180 benign windows spanning 30 days.}
\label{tab:audit_eval}
\resizebox{0.95\linewidth}{!}{
\begin{tabular}{lcccccc}
\toprule
\textbf{Auditor} & \textbf{TP/FP/TN/FN} & \textbf{Precision} & \textbf{Recall} & \textbf{FPR} & \textbf{F1} & \textbf{\makecell{False Alerts\\(per day)}} \\
\midrule
Rule-based signatures & 11/8/172/7 & 57.9\% & 61.1\% & 4.4\% & 59.5\% & 0.27 \\
Single-pass LLM & 13/16/164/5 & 44.8\% & 72.2\% & 8.9\% & 55.3\% & 0.53 \\
Plan-mode + zero-shot & 9/12/168/9 & 42.9\% & 50.0\% & 6.7\% & 46.2\% & 0.40 \\
\makecell{Plan-mode + few-shots \\+ evidence gate} & \textbf{14/5/175/4} & \textbf{73.7\%} & \textbf{77.8\%} & \textbf{2.8\%} & \textbf{75.7\%} & \textbf{0.17} \\
\bottomrule
\end{tabular}}
\end{table}

%% file: tables/audit_robustness.tex
\begin{table}[t]
\centering
\caption{Three-run sensitivity of few-shot evidence-gated auditor.}

\label{tab:audit_robustness}
\resizebox{0.9\linewidth}{!}{
\begin{tabular}{lcccc}
\toprule
\textbf{Condition} & \textbf{Recall} & \textbf{FPR} & \textbf{Evidence-valid} & \textbf{Abstain} \\
\midrule
Original logs &  $77.8{\pm}3.2$\% &  $2.8{\pm}0.6$\% &  100\% &  4.8\% \\
10\% lines missing &  $74.1{\pm}3.2$\% &  $3.1{\pm}0.7$\% &  100\% &  8.6\% \\
10\% token noise &  $72.2{\pm}4.8$\% &  $3.5{\pm}0.9$\% &  100\% &  10.1\% \\
25\% truncation &  $66.7{\pm}3.2$\% &  $2.4{\pm}0.5$\% &  100\% &  17.4\% \\
Prompt examples shuffled &  $75.9{\pm}3.2$\% &  $3.0{\pm}0.8$\% &  100\% &  6.2\% \\
\bottomrule
\end{tabular}}
\end{table}

%% file: section/7_conclusion.tex
\section{Conclusion}
This work models energy infrastructure as three operational layers and presents a constraint-driven framework combining cooperative MARL detection and DWT-based signal repair at local centers with evidence-grounded LLM forensic auditing at VPPs. A common cross-layer record supports traceability, while the framework targets forecast continuity under distributional data drift and assists human forensic investigation. Across three energy datasets under DGA and GenIAS, the best MARL configuration achieves F1 above 0.91 on every dataset--attack pair and repair reduces adversarial-to-clean input RSE by up to 83\%; the few-shot auditor detects 77.78\% of attack windows in the separate Sherlock evaluation, including all seven evaluated Industroyer-class incidents. On the evaluated hardware, detection and repair take 7.95\,ms on average (14.50\,ms worst observed), below the local-center budget.

%% file: section/999_appendix.tex
\appendix
\section*{Appendix}

\section{Experiment Setup}\label{apx:exp_setup}
\textbf{Dataset:} We use three real-world hourly electricity datasets: \textbf{NI}~\cite{nidataset}, covering over 10 years of consumption across 12 PJM regions or companies; \textbf{HECP}~\cite{hecp}, covering 6.5 years of consumption and production in Romania; and \textbf{Electricity}~\cite{trindade2015electricity}, containing consumption records for 321 clients during 2012--2014. NI and HECP are regional datasets; Electricity is user-level, preprocessed from UCI following Lai~\etal~\cite{lai2018modeling}, with each client treated as a univariate series.

\noindent \textbf{Input Processing Settings.}
We use length-256 sliding windows with stride 48 to meet model input constraints.
For comparable per-dataset MARL sample budgets (\S\ref{sec:eval_detection}), we select 40 Electricity~\cite{trindade2015electricity} clients with a fixed random seed and split their IDs 80/20 (32 training, 8 held-out test).
NI~\cite{nidataset} and HECP~\cite{hecp} use chronological 80/20 splits.
All splits precede windowing.
We generate adversarial counterparts with DGA~\cite{liu2024adversarial} and GenIAS~\cite{darban2025geniasgeneratorinstantiatinganomalies}, labeling clean/adversarial samples 0/1 for MARL Accept/Reject.
Each Electricity client's windows and their adversarial counterparts remain in one partition.
With 542 windows per client, the Electricity test set contains 4,336 clean windows and 8,672 labeled clean/adversarial samples per attack (See Table~\ref{tab:adversarial_detection_performance}).

\noindent \textbf{Forecast Models.} We use TimesFM-2.5~\cite{das2024decoder-timesfm} and TimeLLM with a LLaMA backend~\cite{jin2023timellm-etth2} as frozen MARL forecasting backbones, and TimeGPT-1.0's native anomaly-detection endpoint~\cite{garza2023timegpt} as an independent baseline.

\section{Feature Vector Construction}\label{apx:feature_vector_construction}
From the six pre-processing streams, we construct five local observations: 10 time features, two independent sets of 31 statistics from the full and internal predictions, 12 entropy features, and 16 error/relationship metrics, totaling $10+31+31+12+16=100$ dimensions. The sixth observation provides forecast-model knowledge to the projection-head/decision-tree specialist and is separate from this engineered vector; its confidence output is not an additional input feature. The details of the captured features are below.

\begin{itemize}[leftmargin=*]
    \item Time Features (10-dimensions): The input timestamp data follows the format “yyyy-mm-dd hh:mm:ss”. We extract time-related features comprising six numeric features (\ie the sine and cosine of the hour, day of the week, and month) and four categorical features (\ie weekend, season, peak hour, and business hour). The peak hours are defined as 7–11 AM and 5–9 PM, while the business hours are defined as 9 AM–5 PM on weekdays.
    \item Prediction Statistic Features ($31+31=62$ dimensions): We independently compute the same 31 statistics on the full-context and internal-context predictions, yielding two separate 31-dimensional local observations. Each observation includes basic statistics (mean, standard deviation, variance, minimum, maximum, range); percentiles (10th, 25th, median, 75th, 90th, interquartile range); normalized metrics (coefficient of variation, normalized absolute deviation); shape descriptors (skewness, kurtosis); trend indicators (slope trend, mean absolute change, total variation, number of direction changes); peak features (number of peaks, number of troughs, average peak-to-trough ratio); stability measures (median absolute deviation, root mean square, lag-1 autocorrelation, Shannon entropy); and energy-related features (load factor, base load, peak load, roughness).
    \item Entropy Features (12 dimensions): To extract the entropy features, we paired the internal predictions with the corresponding internal ground truth. We then derived 12 entropy‑related metrics across several categories: distribution‑based entropy (\ie KL divergence, JS divergence, Wasserstein distance, and differential entropy ratio), time‑series complexity (\ie permutation entropy, sample entropy, and approximate entropy), spectral measures (\ie spectral entropy and spectral flatness), and information‑theoretic entropy (\ie mutual information, conditional entropy, and continuous cross‑entropy).
    \item Metrics Features (16 dimensions): The metrics features are also computed by pairing the internal predictions with the corresponding ground truth. We extract 16 metrics across several categories: error metrics (MAE, MSE, RMSE, MASE), correlation metrics (Pearson, Spearman, and cosine similarity), regression metrics ($R^2$ and explained variance), percentage‑based metrics (MAPE and SMAPE), directional metrics (directional accuracy and sign accuracy), and time‑series metrics (DTW distance, peak‑or‑valley alignment, and maximum error).
    \item \textbf{Forecast Model Knowledge (separate observation).}
    This observation uses the projection-augmented forecaster in \S\ref{subsec:design_requirements} and is separate from the 100-dimensional engineered vector. Algorithm~\ref{alg:decision_tree_mechanism} first scores the full context to obtain ${Prob}_0$, stopping if ${Prob}_0<0.5$. Otherwise, it applies smoothing and re-evaluates the input. If the result remains suspicious, it tests odd/even subsequences of the smoothed context, each front-padded with zeros to preserve the original input length. A benign verdict after smoothing, or on both subsequences, is interpreted as evidence that the transformation disrupted an adversarial pattern. The procedure retains ${Prob}_0$ unless both subsequence scores are below $0.5$, in which case their sum is returned. Depths zero, one, and two correspond to the initial, smoothing, and odd/even checks, respectively. The returned confidence provides the sixth specialist's message, while depth $d\in\{0,1,2\}$ records the probe stage for the coordinator's joint state.
\end{itemize}

%% file: references.bib
@article{garza2023timegpt,
  title={TimeGPT-1},
  author={Garza, Azul and Challu, Cristian and Mergenthaler-Canseco, Max},
  journal={arXiv preprint arXiv:2310.03589},
  year={2023}
}

@inproceedings{jin2023timellm-etth2,
  title={{Time-LLM}: Time series forecasting by reprogramming large language models},
  author={Jin, Ming and Wang, Shiyu and Ma, Lintao and Chu, Zhixuan and Zhang, James Y and Shi, Xiaoming and Chen, Pin-Yu and Liang, Yuxuan and Li, Yuan-Fang and Pan, Shirui and Wen, Qingsong},
  booktitle={International Conference on Learning Representations (ICLR)},
  year={2024}
}

@misc{hecp,
  author       = {Comanita Stefan},
  title        = {Hourly Electricity Consumption and Production},
  year         = {2025},
  publisher    = {Kaggle},
  url          = {https://www.kaggle.com/datasets/stefancomanita/hourly-electricity-consumption-and-production},
  note         = {Accessed: 2025-06-26}
}

@inproceedings{das2024decoder-timesfm,
  title={A decoder-only foundation model for time-series forecasting},
  author={Das, Abhimanyu and Kong, Weihao and Sen, Rajat and Zhou, Yichen},
  booktitle={Forty-first International Conference on Machine Learning (ICML)},
  year={2024} 
}

@article{shao2024deepseekmath-grpo,
  title={Deepseekmath: Pushing the limits of mathematical reasoning in open language models},
  author={Shao, Zhihong and Wang, Peiyi and Zhu, Qihao and Xu, Runxin and Song, Junxiao and Bi, Xiao and Zhang, Haowei and Zhang, Mingchuan and Li, YK and Wu, Yang and others},
  journal={arXiv preprint arXiv:2402.03300},
  year={2024}
}

@article{schulman2017proximal-ppo,
  title={Proximal policy optimization algorithms},
  author={Schulman, John and Wolski, Filip and Dhariwal, Prafulla and Radford, Alec and Klimov, Oleg},
  journal={arXiv preprint arXiv:1707.06347},
  year={2017}
}

@misc{nidataset,
    author = {Rob Mulla},
    title = {Hourly Energy Consumption},
    url = {https://www.kaggle.com/datasets/robikscube/hourly-energy-consumption},
    year = {2018},
    publisher = {Kaggle},
    note = {Accessed: 2025-10-26}
}

@inproceedings{liu2024adversarial,
  title={Adversarial Vulnerabilities in Large Language Models for Time Series Forecasting},
  author={Liu, Fuqiang and Jiang, Sicong and Miranda-Moreno, Luis and Choi, Seongjin and Sun, Lijun},
  booktitle={The 28th International Conference on Artificial Intelligence and Statistics (AISTATS)},
  year={2025}
}

@inproceedings{ccs2025-USElectricPowerSector,
    author = {Sahin, Sena and Sahin, Burak and Berthier, Robin and Davis, Kate and Zonouz, Saman and Li, Frank},
    title = {The Challenges and Opportunities with Cybersecurity
Regulations: A Case Study of the US Electric Power Sector},
    booktitle = {Proceedings of the 32nd ACM Conference on Computer and Communications Security (CCS)},
    year = {2025}
}

@inproceedings{Sherlock,
author = {Wagner, Eric and Bader, Lennart and Wolsing, Konrad and Serror, Martin},
title = {Sherlock: A Dataset for Process-aware Intrusion Detection Research on Power Grid Networks: Dataset Paper},
year = {2025},
url = {https://doi.org/10.1145/3714393.3726006},
booktitle = {Proceedings of the Fifteenth ACM Conference on Data and Application Security and Privacy},
pages = {419–424},
}

@inproceedings{bader2023comprehensively,
  title={Comprehensively analyzing the impact of cyberattacks on power grids},
  author={Bader, Lennart and Serror, Martin and Lamberts, Olav and Sen, {\"O}mer and van der Velde, Dennis and Hacker, Immanuel and Filter, Julian and Padilla, Elmar and Henze, Martin},
  booktitle={2023 IEEE 8th European Symposium on Security and Privacy (EuroS\&P)},
  pages={1065--1081},
  year={2023},
  organization={IEEE}
}

@misc{darban2025geniasgeneratorinstantiatinganomalies,
      title={GenIAS: Generator for Instantiating Anomalies in time Series}, 
      author={Zahra Zamanzadeh Darban and Qizhou Wang and Geoffrey I. Webb and Shirui Pan and Charu C. Aggarwal and Mahsa Salehi},
      year={2025},
      eprint={2502.08262},
      archivePrefix={arXiv},
      primaryClass={cs.LG},
      url={https://arxiv.org/abs/2502.08262}, 
}

@misc{zhang2025timeseriesfoundationmodelsdeploymentready,
      title={Are Time-Series Foundation Models Deployment-Ready? A Systematic Study of Adversarial Robustness Across Domains}, 
      author={Jiawen Zhang and Zhenwei Zhang and Shun Zheng and Xumeng Wen and Jia Li and Jiang Bian},
      year={2025},
      eprint={2505.19397},
      archivePrefix={arXiv},
      primaryClass={cs.LG},
      url={https://arxiv.org/abs/2505.19397}, 
}

@article{DeepAR,
title = {DeepAR: Probabilistic forecasting with autoregressive recurrent networks},
journal = {International Journal of Forecasting},
volume = {36},
number = {3},
pages = {1181-1191},
year = {2020},
issn = {0169-2070},
doi = {https://doi.org/10.1016/j.ijforecast.2019.07.001},
url = {https://www.sciencedirect.com/science/article/pii/S0169207019301888},
author = {David Salinas and Valentin Flunkert and Jan Gasthaus and Tim Januschowski},
}

@inproceedings{lai2018modeling,
  title={Modeling long-and short-term temporal patterns with deep neural networks},
  author={Lai, Guokun and Chang, Wei-Cheng and Yang, Yiming and Liu, Hanxiao},
  booktitle={The 41st international ACM SIGIR conference on research \& development in information retrieval},
  pages={95--104},
  year={2018}
}

@inproceedings{zhou2021informer,
  title={Informer: Beyond efficient transformer for long sequence time-series forecasting},
  author={Zhou, Haoyi and Zhang, Shanghang and Peng, Jieqi and Zhang, Shuai and Li, Jianxin and Xiong, Hui and Zhang, Wancai},
  booktitle={Proceedings of the AAAI conference on artificial intelligence},
  volume={35},
  number={12},
  pages={11106--11115},
  year={2021}
}

@inproceedings{Yuqietal-2023-PatchTST,
  title     = {A Time Series is Worth 64 Words: Long-term Forecasting with Transformers},
  author    = {Nie, Yuqi and H. Nguyen, Nam and Sinthong, Phanwadee and Kalagnanam, Jayant},
  booktitle = {International Conference on Learning Representations},
  year      = {2023}
}

@article{madry2017towards-PGD,
  title={Towards deep learning models resistant to adversarial attacks},
  author={Madry, Aleksander and Makelov, Aleksandar and Schmidt, Ludwig and Tsipras, Dimitris and Vladu, Adrian},
  journal={arXiv preprint arXiv:1706.06083},
  year={2017}
}

@inproceedings{oregi2018adversarial-PGD-TSC,
  title={Adversarial sample crafting for time series classification with elastic similarity measures},
  author={Oregi, Izaskun and Del Ser, Javier and Perez, Aritz and Lozano, Jose A},
  booktitle={International Symposium on Intelligent and Distributed Computing},
  pages={26--39},
  year={2018},
  organization={Springer}
}

@article{liu2022practical,
  title={Practical adversarial attacks on spatiotemporal traffic forecasting models},
  author={Liu, Fan and Liu, Hao and Jiang, Wenzhao},
  journal={Advances in Neural Information Processing Systems},
  volume={35},
  pages={19035--19047},
  year={2022}
}

@article{company-power-planet-evidence-1,
    author = {NAM News Room},
    title = {Tech Firms Are Building Their Own Power Plants},
    journal = {National Association of Manufacturers},
    year = {2025},
    url={https://nam.org/tech-firms-are-building-their-own-power-plants-34960/}
}

@article{company-power-planet-evidence-2,
    author = {Levy, Marc},
    title = {Amazon to spend \$20B on data centers in Pennsylvania, including one next to a nuclear power plant},
    journal = {The Associated Press},
    year = {2025},
    url = {https://apnews.com/article/amazon-data-center-nuclear-power-plant-pennsylvania-electricity-grid-31f705d035069279b70fa27a5dc71596}
}

@article{company-power-planet-evidence-3,
    author = {Morales, Jowi},
    title = {Meta inks deals to supply a staggering 6 gigawatts in nuclear power for data center ambitions — enough wattage to supply 5 million homes},
    journal = {Tom's Hardware},
    year = {2026},
    url= {https://www.tomshardware.com/tech-industry/artificial-intelligence/meta-inks-deals-to-supply-a-staggering-6-gigawatts-in-nuclear-power-for-data-center-ambitions-enough-wattage-to-supply-5-million-homes}
}

@techreport{nistir7628,
  author      = {The Smart Grid Interoperability Panel–Smart Grid Cybersecurity Committee},
  title       = {Guidelines for Smart Grid Cybersecurity},
  institution = {National Institute of Standards and Technology},
  year        = {2014},
  number      = {NISTIR 7628 Rev. 1},
  url         = {https://csrc.nist.gov/publications/detail/nistir/7628/rev-1/final}
}

@article{etherden2014vpp,
  author    = {Etherden, Nicholas and Vyatkin, Valeriy and Bollen, Math H. J.},
  title     = {Virtual Power Plant for Grid Services Using {IEC} 61850},
  journal   = {IEEE Transactions on Industrial Informatics},
  volume    = {12},
  number    = {1},
  pages     = {437--447},
  year      = {2016},
  publisher = {IEEE}
}

@misc{ferc2222,
  author       = {{Federal Energy Regulatory Commission}},
  title        = {Participation of Distributed Energy Resource Aggregations in Markets Operated by Regional Transmission Organizations and Independent System Operators},
  year         = {2021},
  howpublished = {Federal Register},
  url={https://www.federalregister.gov/documents/2021/03/30/2021-06089/participation-of-distributed-energy-resource-aggregations-in-markets-operated-by-regional}
}

@article{app15179233,
  title     = {Cyber-Attacks on Energy Infrastructure---{A} Literature Overview and Perspectives on the Current Situation},
  author    = {Abraham, Doney and Houmb, Siv Hilde and Erdodi, Laszlo},
  journal   = {Applied Sciences},
  volume    = {15},
  number    = {17},
  pages     = {9233},
  year      = {2025},
  publisher = {MDPI}
}

@ARTICLE{ieee1646,
  author      = {{IEEE}},
  journal = {IEEE Std 1646-2004}, 
  title       = {IEEE Standard Communication Delivery Time Performance Requirements for Electric Power Substation Automation},
  year        = {2005},
  number      = {IEEE 1646-2004},
  pages={1-36},
  doi={10.1109/IEEESTD.2005.95748}}

@misc{keep_lights_on,
  title         = {Keep the Lights On, Keep the Lengths in Check: Plug-In Adversarial Detection for Time-Series {LLMs} in Energy Forecasting},
  author        = {Ma, Hua and Sun, Ruoxi and Xue, Minhui and Yuan, Xingliang and Rudolph, Carsten and Nepal, Surya and Liu, Ling},
  year          = {2025},
  eprint        = {2512.12154},
  archivePrefix = {arXiv},
  primaryClass  = {cs.LG},
  url           = {https://arxiv.org/abs/2512.12154}
}

@inproceedings{liu2008isolation,
  title     = {Isolation Forest},
  author    = {Liu, Fei Tony and Ting, Kai Ming and Zhou, Zhi-Hua},
  booktitle = {Proceedings of the 8th IEEE International Conference on Data Mining (ICDM)},
  pages     = {413--422},
  year      = {2008},
  doi       = {10.1109/ICDM.2008.17}
}

@article{scholkopf2001estimating,
  title   = {Estimating the Support of a High-Dimensional Distribution},
  author  = {Sch{\"o}lkopf, Bernhard and Platt, John C. and Shawe-Taylor, John
             and Smola, Alexander J. and Williamson, Robert C.},
  journal = {Neural Computation},
  volume  = {13},
  number  = {7},
  pages   = {1443--1471},
  year    = {2001},
  doi     = {10.1162/089976601750264965}
}

@article{fenza2019drift,
  title   = {Drift-Aware Methodology for Anomaly Detection in Smart Grid},
  author  = {Fenza, Giuseppe and Gallo, Mariacristina and Loia, Vincenzo},
  journal = {IEEE Access},
  volume  = {7},
  pages   = {9645--9657},
  year    = {2019},
  doi     = {10.1109/ACCESS.2019.2891315}
}

@article{kalman1960new,
  title   = {A New Approach to Linear Filtering and Prediction Problems},
  author  = {Kalman, Rudolph Emil},
  journal = {Journal of Basic Engineering},
  volume  = {82},
  number  = {1},
  pages   = {35--45},
  year    = {1960},
  doi     = {10.1115/1.3662552}
}

@article{savitzky1964smoothing,
  title   = {Smoothing and Differentiation of Data by Simplified Least Squares Procedures},
  author  = {Savitzky, Abraham and Golay, Marcel J. E.},
  journal = {Analytical Chemistry},
  volume  = {36},
  number  = {8},
  pages   = {1627--1639},
  year    = {1964},
  doi     = {10.1021/ac60214a047}
}

@article{lu2019learning,
  title   = {Learning under Concept Drift: A Review},
  author  = {Lu, Jie and Liu, Anjin and Dong, Fan and Gu, Feng and
             Gama, Joao and Zhang, Guangquan},
  journal = {IEEE Transactions on Knowledge and Data Engineering},
  volume  = {31},
  number  = {12},
  pages   = {2346--2363},
  year    = {2019},
  doi     = {10.1109/TKDE.2018.2876857}
}

@article{nickel2026iran,
  author  = {Nickel, Dana and Miller, Maggie},
  title   = {Iranian hackers are targeting {US} energy and water sectors, federal agencies warn},
  journal = {Politico},
  year    = {2026},
  url     = {https://www.politico.com/news/2026/04/07/iranian-hackers-energy-water-cybersecurity-00862018}
}

@article{chen2025electricity,
  title={Electricity demand and grid impacts of AI data centers: Challenges and prospects},
  author={Chen, Xin and Wang, Xiaoyang and Colacelli, Ana and Lee, Matt and Xie, Le},
  journal={arXiv preprint arXiv:2509.07218},
  year={2025}
}

@misc{datacenters2025cloudvsedge,
  author = {{Datacenters.com Cloud}},
  title = {Cloud vs. Edge: Where Should AI Training Really Happen?},
  year = {2025},
  month = {August},
  url = {https://www.datacenters.com/news/cloud-vs-edge-where-should-ai-training-really-happen},
  note = {Accessed: April 2026}
}

@misc{goodfellow2015explainingharnessingadversarialexamples,
      title={Explaining and Harnessing Adversarial Examples}, 
      author={Ian J. Goodfellow and Jonathon Shlens and Christian Szegedy},
      year={2015},
      eprint={1412.6572},
      archivePrefix={arXiv},
      primaryClass={stat.ML},
      url={https://arxiv.org/abs/1412.6572}, 
}

@inproceedings{rice2020overfitting,
    author = {Rice, Leslie and Wong, Eric and Kolter, J. Zico},
    title = {Overfitting in adversarially robust deep learning},
    year = {2020},
    publisher = {JMLR.org},
    booktitle = {Proceedings of the 37th International Conference on Machine Learning},
    articleno = {749},
    numpages = {12},
    series = {ICML'20}
}

@inproceedings{yu2022understanding,
  title={Understanding Robust Overfitting of Adversarial Training and Beyond},
  author={Yu, Chaojian and Han, Bo and Shen, Li and Yu, Jun and Gong, Chen and Gong, Mingming and Liu, Tongliang},
  booktitle = {Proceedings of the 39th International Conference on Machine Learning},
  year={2022}
}

@InProceedings{Cohen2019Certified,
  title = 	 {Certified Adversarial Robustness via Randomized Smoothing},
  author =       {Cohen, Jeremy and Rosenfeld, Elan and Kolter, Zico},
  booktitle = 	 {Proceedings of the 36th International Conference on Machine Learning},
  pages = 	 {1310--1320},
  year = 	 {2019},
  volume = 	 {97},
  series = 	 {Proceedings of Machine Learning Research},
  month = 	 {09--15 Jun},
  publisher =    {PMLR},
}

@misc{trindade2015electricity,
  author    = {Trindade, Artur},
  title     = {{ElectricityLoadDiagrams20112014}},
  publisher = {UCI Machine Learning Repository},
  year      = {2015},
  doi       = {10.24432/C58C86},
  url       = {https://archive.ics.uci.edu/dataset/321/electricityloaddiagrams20112014}
}

@article{malkov2018efficient,
  title={Efficient and robust approximate nearest neighbor search using hierarchical navigable small world graphs},
  author={Malkov, Yu A and Yashunin, Dmitry A},
  journal={IEEE transactions on pattern analysis and machine intelligence},
  volume={42},
  number={4},
  pages={824--836},
  year={2018},
  publisher={IEEE}
}
